# Thermoelectric effects and the asymmetry of the current-voltage characteristics of metallic point contacts

### Yu. G. Naidyuk, N. N. Gribov, O. I. Shklyarevskii, A. G. M. Jansen,[1)] and I. K. Yanson

*PhysicoTechnical Institute of Low Temperatures, Academy of Sciences of the Ukrainian SSR, Khar'kov*

An asymmetry as a function of the direction of current flow is observed in the current-voltage characteristic (CVC) and its first and second derivatives for point heterocontacts between pure metals {Cu, Ni, Fe) as well as between these metals and dilute alloys (CuFe, CuMn). It is shown that the observed asymmetry is caused by thermoelectrical phenomena (Seebeck, Peltier, and Thompson effects), observed when the temperature inside the contact differs from the temperature of the bath. In the low energy range (less than or of the order of the Debye energy) the asymmetry of the CVC is affected mainly by the Seebeck effect, while at high energies and for larger contacts (lower resistance) the contributions of all noted effects are of the same order of magnitude. A technique is proposed for determining the temperature of a heterocontact by measuring the CVC in two polarities. It is established that in the intermediate (between the diffusion and ballistic) state of flight of the electrons through the constriction the temperature of the heterocontact increases linearly with the voltage on it, and all the more rapidly the larger the contact and the more impurity in it.

The current-voltage characteristic (CVC) of a point contact of size d, much less than the energy relaxation length ($l_e$) and momentum relaxation length ($l_i$) of the electrons, at helium temperatures ($T \ll \Theta_D$, $\Theta_D$ is the Debye temperature) contains spectroscopic information about the mechanisms of scattering of conduction electrons.[1,2] Thus, for example, the second derivative of the CVC ($d^2V/dI^2$) of such contacts [or the point-contact (PC) spectrum] are directly related with the PC by the electron—phonon interaction (EPI) function $G(eV)$[3]:

$$d^2V/dI^2(eV) \approx -d^2I/dV^2(eV) = aG(eV), \quad (1)$$

where a is some energy-independent constant for a given metal; V and I are the voltage and the current in the contact; and, e ig the electron charge. A relation similar to (1) also holds in the so-called diffusion limit, when $l_i \ll d \ll (l_i l_e)^{1/2}$. In this case the right side of (1) must be multiplied by ($l_i/d$), which only decreases the intensity of the PC spectrum. If all electron mean-free paths becomes less than the contact dimensions ($l_i, l_e \ll d$), then the thermal regime of current flow, in which spectral features are absent in the second derivatives and the temperature inside the contact $T_C$ is related to the voltage on the contact by the relation[4]

$$T_C^2 = T_0^2 + V^2/4L, \quad (2)$$

where $T_0$ is the temperature of the bath and L is Lorentz's number, is realized.

According to the theory,[3,4] the PC spectrum and correspondingly the dependence of the differential resistance of the contact $R(V) = dV/dI(V)$ are symmetrical relative to the polarity of the applied voltage V, in the case of both homo- and heterocontacts.[5] However, in experiments on contacts between a normal metal and a compound with a variable valence[6,7] or a Kondo lattice[8] asymmetry was observed in the dependence $R_d(V)$ relative to the direction of current flow through the contact. In Refs. 7 and 8 it was pointed out that the asymmetry of $R_d(V)$ could be caused by the Seebeck effect in the thermal state of current flow through a heterocontact containing a compound with a variable valence.

In this work the asymmetry of spectra are observed for the first time in the simplest systems, where the electrodes consisted of pure metals (Cu, Ni, Fe) or dilute magnetic alloys (CuFe and CuMn). Asymmetry was observed only in the case of a point contact between different materials. The observed asymmetry is linked with the manifestation of thermoelectric effects, when the temperature inside the contact differs from the temperature at the edges. In addition, in Cu – CuFe contacts, in which the current flows in an intermediate state (between the diffusion and ballistic states), the temperature also differs from the temperature of the bath and depends on the impurity concentration and the diameter of the contact.

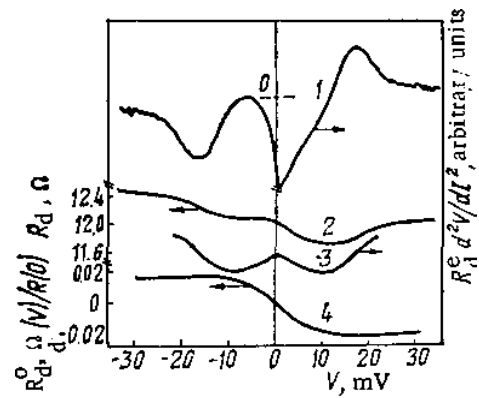

FIG. 1. $d^2V/dI^2(V)$ (1) and $R_d(V)$ (2), as well as the even (3) and odd (4) components of $R_d(V)$ for the heterocontact Cu-CuFe (1 at.%) at $T_0 = 4.2\ K$. For $V > 0$ the current flows from the alloy into the copper.

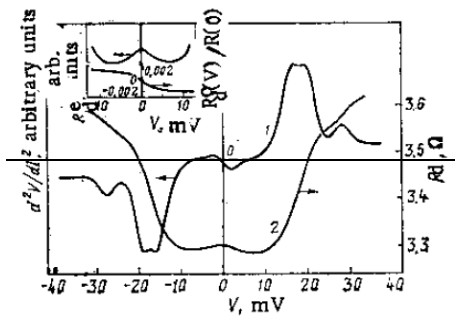

FIG. 2. $d^2V/dI^2(V)$ (1) and $R_d(V)$ (2) for the heterocontact Cu-CuFe (0.01 at. %) at $T_0 = 4.2$ K. The even and odd components of $R_d(V)$ are shown in the inset. For $V > 0$ the current flows from the alloy into the copper.

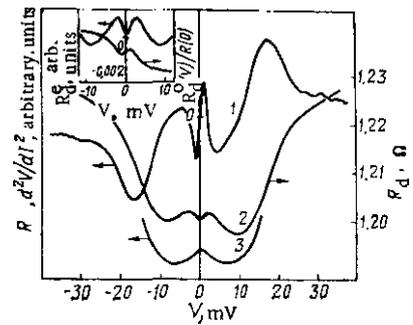

FIG. 3. $d^2V/dI^2(V)$ (1) and $R_d(V)$ (2) for the heterocontact Cu-CuMn (0.6 at. %) and $R_d(V)$ for the heterocontact Cu-CuMn (0.12 at. %) (3) at $T_0 = 4.2$ K. The even and odd components of $R_d(V)$ are shown in the inset. For $V > 0$ the current flows from the alloy into the copper.

## EXPERIMENTAL RESULTS

We shall consider briefly the experimental procedure, which is described in detail in Refs. 1 and 4. The derivatives of the CVC were recorded by the method of synchronous detection of the first $V_1$ and second $V_2$ harmonics of the modulating signal $dV/dI$, $V_2 \sim d^2V/dI^2$). A bridge scheme was used to obtain a more accurate record of $V_1(V)$ or $R_D(V)$. The point contacts were formed both in the needle-anvil geometry[1] and by the shear technique[9] between metals which were first worked by electrochemical polishing in the corresponding solutions. Most measurements were performed at a temperature of 4,2 K.

Figures 1 and 2 show the curves of $R_D(V)$, $d^2V/dI^2$ for the heterocontacts Cu – CuFe with an iron content of 1 and 0.01%. It is evident that the asymmetry is already manifested in the energy range $eV \ll k\Theta_D$, containing the so-called zero anomaly, associated with the scattering of electrons by localized magnetic moments of impurities (Refs. 1 and 10). At the same time the EPI features on the PC spectra at 18 and 28 mV are practically identical for both polarities. After the intensity of the spectral features of the EPI are measured, by using the results of Ref. 11 it is possible to evaluate the ratio $l_i/d$ for the contacts shown in Figs. 1 and 2. It turned out that for a sample with a concentration of 0.01 at.% Fe $l_i/d \sim 0.7$, while for 1 at.% Fe $l_i/d \sim 0.2$. Thus the state of current flow in the given point contacts is an intermediate state between the ballistic and diffusion states.

The curves $R_d(V)$ can be represented in the form of the even $R_d^e(V)$ and $R_d^0(V)$ dependences: $R_d^e(V) = ½ (R_d^+ + R_d^-)$, $R_d^e(V) = ½ (R_d^+ - R_d^-)$. It is evident from Figs. 1 and 2 that the even dependence is similar to $R_d(V)$ for contacts between identical electrodes made of CuFe (Ref. 10), while the odd dependence has the form of the smeared step in the region of $V = 0$.

A qualitatively similar picture is also preserved for heterocontacts with Fe concentrations of 0.14, 0.4, and 1.4 at.%. When the iron concentration is raised the amplitude of the anomaly in $d^2V/dI^2$ at $V=0$ increases above the phonon features, which in this case become smeared.

For Cu – CuMn contacts, an asymmetry was also observed in $R_d(V)$ and the PC spectra (Fig. 3), but for analogous concentrations of Mn it is appreciably smaller than in CuFe contacts [$R_d^0/R_d(0)$ is smaller by almost an order of magnitude]. In addition, $R_d^0(V)$ in Cu – CuMn exhibits a more complicated behavior than $R_d^0,(V)$ for Cu – CuFe, while $R_d^e(V)$ is similar to the dependence $R_d(V)$ in the case of identical CuMn electrodes.[10] We note that the asymmetry in Cu – CuMn heterocontacts was observed for alloys with an Mn concentration of 0.6 and 1 at.%, when the alloy is in a spin-glass state (this is indicated by the minimum on the curve $R_d(V)$ at $V = 0$), while for Mn concentrations of 0.12 at.% and less it has a virtually symmetrical maximum at $V = 0$ (Fig. 3, curve 3).

Figure 4 shows $R_d(V)$ and the CVC of a point contact between pure Cu and Fe metals, where an asymmetry is also observed, and $R_d^0(V)$ increases with the voltage on the contact. Qualitatively similar curves were obtained for Cu – Ni, Fe – Ni, Fe – CuFe (1 atm %) heterocontacts, where one of the electrodes was a ferromagnetic metal. In the case of the Ni – CuFe heterocontact (1 at.%) (Fig. 5) the asymmetric CVC, recorded in two polarities, intersect at $V = 125$ mV, and the sign of the dependence $R_d^0(V)$ changes at $V \approx 65$ mV. Figure 6 shows the PC spectra ($d^2V/dI^2$) of the Cu – Ni heterocontact, in which a sharp maximum is observed at $V \approx 200$ mV. It is evident that the right boundary of this maximum intersects the $d^2V/dI^2 = 0$ axis at different voltages depending on the polarity. Thus the features in $d^2V/dI^2(V)$ are shifted relative to one another by $\Delta V = V^+ - V^- \approx 10$ mV, i.e., asymmetry is also observed in their position.

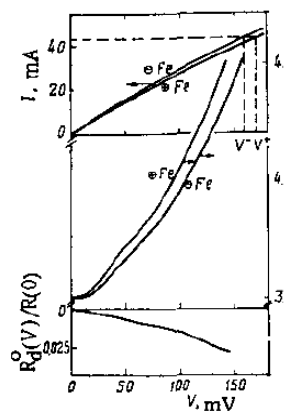

FIG. 4. CVC and $R_d(V)$ for a Cu-Fe heterocontact in two polarities and the odd part $R_d^0(V)$ at $T_0 = 4.2$ K.

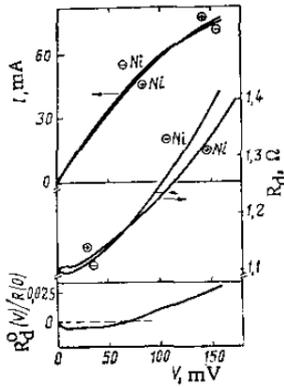

FIG. 5. CVC and $R_d(V)$ for the heterocontact Ni-CuFe (1 at.%) in two polarities and the odd part $R_d^0(V)$ at $T_0 = 4.2$ K.

DISCUSSION

We shall begin our discussion with the results presented last. According to Ref. 4, in contacts consisting of ferromagnetic metals at $eV \gg k\Theta_D$, the thermal state of current flow is realized and the temperature of the contact is determined by the voltage on it in accordance with the formula (2). At a voltage such that the temperature of the contact is equal to the Curie temperature $\Theta_C$, a sharp feature appears on the curve $d^2V/dI^2(V)$ for Ni at $V \approx 200$ mV (Ref. 4). An anomaly of this type is observed in the case of the Cu – Ni heterocontact (Fig. 6), which is a direct indication of the fact that the thermal state of current flow is realized in it. According to Refs. 7 and 8, the asymmetry of $R_d(V)$ and IVC in the thermal state could be associated with thermoelectric effects. Taking into account the Seebeck effect, it is easy to show that

$$\Delta V(V) = 2\int_{T_0}^{T_C} [S_A(T) - S_B(T)]dT, \quad (3)$$

where $\Delta V(V)$ is the difference between the IVC recorded in two polarities with the same magnitude of current flowing through the contact (see Fig. 4); $S_A(T)$, $S_B(T)$ is the thermo-emf of the metals forming the contact. The integral thermo-emf of the pair forming the heterocontact, which in one polarity is summed when a voltage is applied to the contact and is subtracted in the other, whence appears the factor of 2 in front of the integral (3), actually appears on the right side of this equation. Once (3) is calculated for the Cu – Ni pair up to the Curie temperature of nickel ($T_C = \Theta_D = 630$ K) we obtain $\Delta V \approx 22$ mV, which is in good agreement with the measured value of $\Delta V$ in Fig. 6. The values of $S(T)$ used for the calculations are shown in Fig. 7. For the contact Ni – CuFe (1 at.%) the IVC intersect (see Fig. 5), i.e., $\Delta V = 0$ and the integral (3) vanishes. This occurs at $T_C \approx 260$ K as a result of the change in sign of the integrand, since, as can be seen from Fig. 7, the curves $S(T)$ for Ni and for CuFe (1 at.%) intersect. Thus with a voltage of $V \approx 125$ mV on the contact the temperature of the contact, according to (3), must be equal to 260 K. This is somewhat less than follows from the theory of the thermal regime, where a voltage of 125 mV corresponds to a temperature of $T \approx 400$ K. This difference can be linked to the deviation of the properties of a real heterocontact from the model of the thermal state.

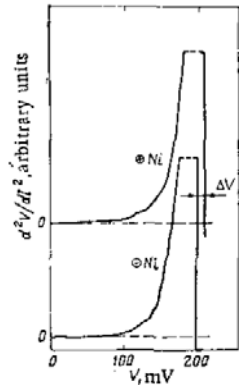

FIG. 6. $d^2V/dI^2(V)$ for the heterocontact Cu-Ni in two polarities at $T_0 = 4.2$ K.

The formula (3) relates the quantities V and $T_C$, so that by using the dependence $\Delta V(V)$ found from the IVC and the value of $S(T)$ for the contact pairs, it is possible to obtain the dependence $T_C(V)$, which is shown in Fig. 8, for three types of heterocontacts. Data for each point were averaged aver the values for five contacts in the case of Cu – Ni, three contacts in the case of Cu – Fe, and five contacts for Fe – Ni. The spread of the values of $T_C$ determined in this manner (shown in Fig. 8b by the "whiskers") constituted several tens of percents, which can be associated with the difference in the contacts both with regard to dimensions and structure, i.e., the ratio between d, $l_\varepsilon$, and $l_i$ could vary in each specific case. It is evident from Fig. 8a that the temperature within the contact at first increases slowly with V, and for $V > 100$ mV the curves $T_C(V)$ are virtually parallel to the theoretical dependence. This behavior of $T_V(V)$ can be explained by the fact that the inelastic mean-free path l at first decreases as V increases owing to the EPI (since $\Theta_D < \Theta_C$), and then because of the electron — magnon scattering and only at energies eV $\sim 100$ mV $> k\Theta_C$ the condition of the thermal state $l_\varepsilon \ll d$ is achieved. Of course, for a more accurate reconstruction of $T_C(V)$ it is necessary to have a theory of the thermal state for heterocontacts which takes into account the thermoelectric phenomena, but, as is evident from the foregoing presentation, the asymmetry of the CVC of point contacts in the thermal state can be satisfactorily explained by the influence of the Seebeck effect.

As noted previously, the heterocontacts Cu – CuFe and Cu – CuMn are in an intermediate state between the diffusion and ballistic states. Let us suppose that the temperature within the contact increases with the voltage on the contact even in this case, and let us try to invoke the Seebeck effect in order to explain the asymmetry of $R_d(V)$ in Cu – CuFe and Cu – CuMn. Since it is difficult to measure $\Delta V$ accurately enough on the CVC of the given heterocontacts, we shall write the

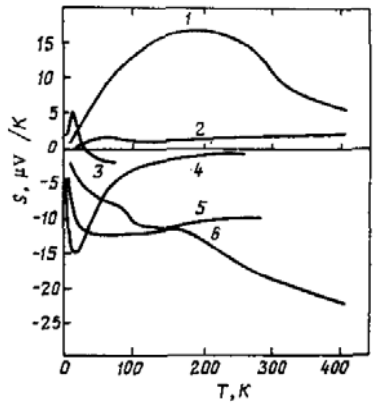

FIG. 7. The curves of the thermo-emf $S(T)$ for Fe (l),[13,15] Cu (2),[12,15] CuMn (0.5 at. %)(3),[16] CuFe (0.01 at. %) (4),[14] CuFe (1 at. %) (5),[14] and Ni (6).[13,15]

formula (3) in the following form:

$$R_d^0(V)/R_{st}^e(V)V = \int_{T_0}^{T_C}\left[S_A(T)-S_B(T)\right]dT, \quad (4)$$

where $R_{st}(V)$ is the static resistance

$$R_{st}(V) = V/I(V) = V/\int_0^V R_{st}^{-1}(\vartheta)d\vartheta;$$
$$R_{st}^0 = 1/2\left(R_{st}^+ - R_{st}^-\right); \quad R_{st}^e = 1/2\left(R_{st}^+ + R_{st}^-\right).$$

The left part of Eq. (4) was determined from the measured dependences $R_d(V)$, and data for $S(T)$ from Fig. 7 were used to calculate the integral. Thus by satisfying the equality (4), as in the preceding case, it is possible to obtain the curve $T_C(V)$ shown in Fig. 8a, It is evident that $T_C(V)$ is well described by a linear function, and the temperature of the contact increases all the more slowly the lower the impurity content and the higher the resistance of the contact. In the heterocontact Cu – CuFe (0.01 at.%) the temperature in the region of phonon energies increases by only 2-3 K, which has virtually no effect on the smearing of the phonon features (see Fig. 2). In the heterocontact Cu – CuFe (1 at.%) the temperature in the contact with R ≈ 12 Ω is equal to 25-40 K, which is sufficient for smearing of the fine features in the region of the transverse phonons (at V = 18 mV) and complete disappearance of the longitudinal peak (at V = 28 mV), At the same time the width and force of the transverse-phonon peak correspond to the width and form of this feature on the PC spectrum of pure copper, measured at 25 K (Ref. 1), which Is an indirect confirmation of the increase in the temperature in the region of the contact.

In the heterocontact Cu – CuMn (0.6 at.%) the curve $R_{st}^0(V)$ [here $R_{st}^0(V) \approx R_d^0(V)$] has a complicated form (see Fig. 3), It s values are greater than zero for 0 – 2 mV, and for V > 2 mV they become negative. Once the integral on the right side of the formula (4) is calculated and the data for the thermo-emf of the alloy CuMn are taken into account (0.4 at.%), we find that the integral is greater than zero in the region from 0 to 55 K. It follows from this that even for V ≤ 2 mV the the temperature inside the contact must be approximately 55 K. This is, however, unrealistic, since in this case there should be virtually no spectral features on the PC spectrum both in the region of the phonon energies and at V = 0 (see Fig. 3). It is more natural to assume that in the given contact S(T) differs from the thermo-emf of a bulk sample. It is evident from Fig. 7, for example, that an insignificant iron impurity can radically change the form of the dependence S(T).

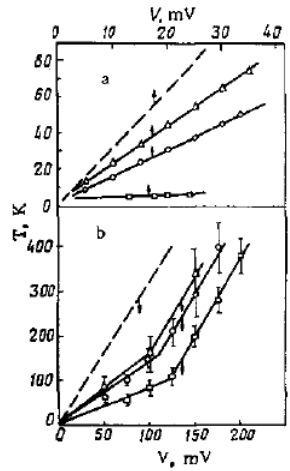

FIG. 8. The temperature in the region of the contact versus the voltage on it for heterocontacts: a) Cu-CuFe (1 at. %), R(0)= 11.6 Ω (O); Cu-CuFe (1 at. %) R(0) = 0.8 Ω (Δ); Cu-CuFe (0.01 at. %), R(0) = 3.3 Ω (□) b) CuFe (Δ), Ni- Fe(O); Cu- Ni (□). The broken line shows the dependence of the temperature on the voltage in the thermal state.[4]

We note that in the thermal state the Peltier effect, which consists of absorption or liberation of heat in the contact accompanying passage of a current through the contact, can contribute to the asymmetry. For one polarity the Peltier heat $W_\Pi^{AB} = (\Pi_A - \Pi_B)I = (S_A - S_B)TI$ will add to the Joule heat, while in the other polarity it will subtract from the Joule heat, so that the temperature of the contact will be somewhat higher in the first case and somewhat lower in the second case than the temperature in the uniform contact. Therefore the resistance of the contact for two polarities will also differ, since it depends on the temperature. Starting from qualitative considerations we find that in this case

$$R_{st}^0(V)/R_{st}^e(V) = \frac{\partial R_{st}(V)}{\partial T}\frac{\left[S_A(T)-S_B(T)\right]}{R_{st}^e(V)V}T^2. \quad (5)$$

The derivative $\partial R_{st}^e(V)/\partial T$ was calculated using the relationship $T_C(V)$, obtained previously. Once $R_{st}^0(V)/R_{st}^e(V)$ is evaluated, and use is made of the formula (5), we find that for Cu – CuFe contacts the contribution of the Pdtier effect to the asymmetry of $R_{st}(V)$ at V ~ 35 mV is only several percent of the contribution of the Seebeck effect. At the same time it increases up to 20% in Cu – Fe and Cu – Ni contacts at V ≈ 150 mV, while in Ni – Fe contacts for V > 150 mV the contributions of both effects to the asymmetry of the CVC are of the same order of magnitude. We call attention to the fact that the contribution of the Peltier effect increases with the voltage (temperature) across the contact and with a decrease in the resistance. This is qualitatively evident from the formula (5), if it is kept in mind that $V \sim T$ in the thermal state.

In order to be systematic, we shall also study the influence of the Thompson effect on the asymmetry of the CVC of the heterocontacts. This effect is manifested in the presence of a temperature gradient and an electric field in the conductor, which, of course, occurs in point contacts. In a nonuniformly heated conductor, passage of a current is accompanied by liberation of absorption of Thompson heat per unit time equal to $w_T = -\mu_T j \nabla T$, where $\mu_T = T(dS/dT)$ is the Thompson coefficient and j is the current density. It is not difficult to show that for a symmetrical distribution of the temperature in the contact of two metals A and B the liberated or absorbed Thompson heat will be equal to $w_T^{AB} = -j \nabla T T(d/dT)[S_A(T) - S_B(T)]$. Assuming that the temperature gradient in the contact and liberation of Peltier heat occur in the same region (of the order of the characteristic size of the contact), we find that $w_T^{AB}/w_\Pi^{AB} \sim -T(d/dT)[S_A(T) - S_B(T)]/[S_A(T) - S_B(T)]$, where $w_\Pi^{AB}$ is the Peltier heat liberated per unit time per unit volume. It is evident from this that the contributions of the effects under study to the asymmetry add if ($S_A - S_B$) and its derivative have different signs and subtract in the opposite case. In addition, if $S_A - S_B \sim -T$, then, $w_T^{AB} \sim w_\Pi^{AB}$, and the Peltier and Thompson effects in $R_d(T)$ will be of the same order of magnitude.

It is evident from the curves S(T) (Fig. 7) that in the case of the Cu – Ni contact the thermo-emf of copper can be neglected, and for nickel $S(T) \sim -T$. Thus for Cu – Ni contacts the effects studied are of the same order of magnitude, though they have different signs, i.e., they can compensate one another. The same can be said about Cu – Fe and Fe – Ni contacts in the region of temperatures up to 200 K, at which S(T) for Fe has a maximum. At temperatures above 200 K the Thompson and Peltier heats in the Fe – Cu contact add, while in the Fe – Ni contact at 200 K < T < 500 K $S_{Fe} - S_{Ni} \approx$ const, i.e., the Tompson heat is equal to zero, and only the Peltier and Seebeck effects contribute to the asymmetry. Thus, in explaining the asymmetry of the CVC of heterocontacts in the thermal state with the help of thermoelectric phenomena, it is necessary to take into account each effect: Seebeck, Peltier, and Thompson.

Summarizing, we note that the asymmetry of the CVC of heterocontacts in the thermal state, observed in this work, can be explained in terms of the thermoelectric effects. The Seebeck effect makes the main contribution to the asymmetry, but with large biases on the contact (eV > k$\Theta_D$) the Peltier and Thompson effects must be kept in mind. In the case of Cu – CuFe contacts it has been established that even in a state intermediate between the diffusion and ballistic states the temperature in the region of the contact differs from the temperature of the bath and depends on the dimensions of the contact and the concentration of the impurity in it. Keeping the latter in mind, we note that the smearing of the phonon features on the PC spectra of CuFe alloys in Ref. 10 as well as other alloys and metals is most likely linked not with the structural disordering of the metal in the contact, but rather with the increase in the temperature in the contact and with the degradation of the resolution of the method, which is proportional to 5.44 kT (Refs. 1 and 2).

In conclusion the authors thank V, M. Kirzhner and V. V. Fisun for assistance in performing the calculations on the Elektronika D3-28 minicomputer.

## NOTATION

Here $\Theta_D$ is the Debye temperature, $\Theta_C$ is the Curie temperature; $T_0$ is the temperature at which the measurement is performed; $T_C$ is the contact temperature; e is the electron charge; $l_i$, $l_e$ are the electron momentum and energy transit lengths; I and V are the current in and the voltage on the contact; d is the diameter of the contact; $R_g(V)$ is the differential resistance of the contact; $R_{st}(V)$ is the static resistance of the contact; $R_d^e(V)$ and $R_d^0(V)$ are the even and odd components of $R_d(V)$; $R_{st}^e(V)$ and $R_{st}^0(V)$ are the even and odd components of $R_{st}(V)$; $V_1(V)$, $V_2(V)$ are the first and second harmonics of the modulating signal; k is Boltzmann's constant; L is the Lorentz constant; S(T) is the absolute thermo-emf, $\Sigma(T)$ is the Peltier coefficient; $\mu_T(T)$ is the Thompson coefficient; j is the current density in the contact; and, W and w are the total and specific heats released in the contact.